\documentclass[12pt]{article}
\makeatletter
\usepackage[margin=72.27pt]{geometry}
\parskip\z@
\usepackage[factor=700,stretch=14,shrink=14,step=1]{microtype}
\usepackage{enumitem}
\setlist{topsep=\smallskipamount,itemsep=\smallskipamount,
  parsep=\z@,partopsep=\z@,beginpenalty=0,midpenalty=0,
  endpenalty=0}
\usepackage{amsmath}
\usepackage{amssymb}
\allowdisplaybreaks
\usepackage{graphicx}
\usepackage{tabularx}
\usepackage{booktabs}
\usepackage{multirow}
\usepackage[colorlinks=true,allcolors=blue,implicit=false]{hyperref}
\usepackage{orcidlink}
\newcolumntype{C}{>{\centering\arraybackslash}X}
\newcolumntype{M}{>{\centering\arraybackslash$\displaystyle}X<{$}}
\usepackage[font=small,labelfont=small,bf]{caption}
\DeclareCaptionLabelFormat{table}{Table #2}

\def\E{\mathbb E}
\def\clap#1{\hb@xt@\z@{\hss#1\hss}}

\def\T{^{{\operator@font T}}}

\def\bigskip{\vskip\bigskipamount}
\def\medskip{\vskip\medskipamount}

\newdimen\@saveparindent
\def\heading#1{\bigskip\medskip
  \noindent\centerline{\textbf{#1}}\par
  \nobreak\bigskip\nobreak
  \@saveparindent\parindent
  \parindent\z@
  \everypar{\parindent\@saveparindent}}
\def\fignum{\the\numexpr\thefigure+1\relax}

\usepackage[authordate,doi=false,footmarkoff,
  maxbibnames=4,minbibnames=1]{biblatex-chicago}

\let\revsdnamedelim\@empty
\bibhang\z@

\begin{document}

{\LARGE\topskip=0.7in
\centerline{The Structure of the U.S. Income Distribution}\smallskip
\centerline{Conrad Kosowsky*\kern0.1em\relax\orcidlink{0009-0007-0479-6746}}
\let\thefootnote\relax\footnotetext{*University of Michigan, Department of Economics and Center for the Study of Complex Systems. Email: coko@umich.edu. Replication and data files for this paper are available \href{https://github.com/ckosowsky/structure-us-incomes}{\underline{here}}. Acknowledgements: Thanks to everyone who provided feedback while I worked on this paper including Marisa Eisenberg, John Leahy, Scott Page, Hilary Zedlitz, and multiple anonymous reviewers.}
\footnotetext{Any views expressed are those of the author and not those of the U.S. Census Bureau. The Census Bureau has reviewed this data product to ensure appropriate access, use, and disclosure avoidance protection of the confidential source data used to produce this product. This research was performed at a Federal Statistical Research Data Center under FSRDC Project Number 2679. (CBDRB-FY24-P2679-R11429 and CBDRB-FY25-P2679-R11880)}
}

\vskip 0.3in

\begin{abstract}
\noindent I show that U.S. incomes follow a one-parameter family of probability distributions over more than fifty years of data. I compare statistical models of income, and I highlight the inverse-gamma distribution as a parsimonious model that matches data particularly well and has straightforward theoretical interpretations. However, despite having relatively few parameters, the inverse-gamma distribution still overfits income data. I establish a linear relationship between parameter estimates, and a one-dimensional model emerges naturally when I exploit this relationship. I conclude with theoretical remarks about the model.

\medskip

\end{abstract}

\vskip 0.1in

\noindent In his eponymous use of power laws to describe the distribution of large incomes over a century ago, Pareto established an enduring empirical regularity in the nature of tail incomes across different societies. Although Pareto distributions fit large incomes remarkably well, they do not accurately describe the body of the income distribution, and since Pareto's discovery, economists have modeled incomes using a large number of probability distributions in what \textcite[][p.\ ix]{kleiber-kotz-2003} call ``an unfortunate lack of coordination among various researchers.'' Consensus on which model best fits the data has not arisen since the publication of \textcite{kleiber-kotz-2003}. My goal in writing this paper is to address this gap in the literature by answering the following two questions: (1) of the statistical models in use now, which fits U.S. income data most closely, and can we do any better? and (2) what economic insights follow from using a parametric representation of the income distribution? I hope that by addressing these questions, this paper will serve as a stepping stone toward a full understanding of how and why the income distribution retains its distinctive appearance.

My main contributions in this paper fall into two categories. First, I make several technical contributions around fitting models to income data, namely incorporating negative incomes into my analysis, a primary focus on parametric description and fit to data, and my use of the shifted inverse-gamma distribution. Much statistical modeling of income has incorporated only positive incomes, perhaps because negative incomes are unexpected or unintuitive. However, Current Population Survey data contains a small number of negative incomes that arise from capital depreciation and business losses, and because negative incomes appear in the data, I include them in my analysis by incorporating a location parameter into models of income. Changing the value of a location parameter translates probability density functions horizontally and allows the support of the distribution to contain negative values. Shifted versions of standard statistical models capture the general features of the income distribution almost perfectly. Comparing candidate models, I find that a shifted inverse-gamma distribution describes the data as well as or better than other probability distributions, and it has the same number or fewer parameters. This parsimonious model is new to the income literature and has certain desirable mathematical properties that other distributions do not.

Second, I contribute to the literature on income modeling by identifying overfitting and adding dimension reduction to my analysis. My approach leads to a one-dimensional model of the U.S. income distribution, which is a new finding. A shifted inverse-gamma distribution has three parameters, but parameter estimates move in sync with one another across years. Detailed examination points to a linear relationship between these parameters and time, so I model the scale and shift parameters as linear functions of the shape parameter and the year. I determine coefficients through a multilinear regression on the shifted inverse-gamma parameter estimates from all years of data, and I use the coefficients to replace the usual scale and shift parameters in the original distribution with linear functions. The resulting one-dimensional model fits data as well as the other models I estimate, which suggests that the income distribution is effectively one-dimensional. My finding of direct time-dependence in income scale and location is new to the literature.

The rest of this paper is organized as follows. Section 1 discusses Current Population Survey income data and the models currently in the literature, and section 2 describes the qualities that we should find in a good parametric model of incomes. Section 3 introduces the inverse-gamma distribution in detail and presents fit, parameter estimates, and evidence of overfitting. In section 4, I compare the inverse-gamma distribution to other probability distributions, and in section 5, I discuss related issues such as inequality, income dynamics, and some implications for future research. Section 6 concludes. In the appendix, I discuss technical details of the estimation procedure. This paper focuses exclusively on income, and for literature on the wealth distribution, see \textcite{benhabib-bisin-2018} and \textcite{benhabib-bisin-luo-2017}. Readers who are primarily interested in what distribution I have settled on should skip to equation~(\ref{CSS_InvG_density}) and table~\ref{CSS_InvG_constants}.

\heading{1. Introduction}

Historically, economists have modeled income in many different ways, including well-known distributions such as gamma \autocite{salem-mount-1974} and log-normal \autocite{aitchison-brown-1957}. See the comprehensive work of \textcite{kleiber-kotz-2003} for a review of historical work on this topic. The authors group various distributions according to whether they arise from Pareto, log-normal, gamma, or beta distributions, and a final chapter details miscelaneous distributions, such as those explored by \textcite{champernowne-1952,champernowne-1953}. \textcite{dagum-1977} argues that most statistical models of income typically arise in one of three ways: (1) by direct comparison with data; (2) as a steady state of a model for income dynamics; or (3) from differential equations that describe some aspect of the income distribution. My work in this paper falls squarely into the first of Dagum's categories.

We can divide more recent literature into three different strains. The first strain has focused on the generalized beta distribution, which is a five-parameter family that nests many common distributions as subfamilies.\footnote{The naming conventions may confuse some readers. The generalized beta distribution has five parameters. The generalized beta type I and generalized beta type II distributions both have four parameters and are special cases of the generalized beta distribution.} \textcite{jenkins-2009} evaluates a variety of inequality measures under a generalized beta type II income distribution, and \textcite{graf-nedyalkova-2014} discuss ways of estimating parameters of the type II distribution, including by matching various inequality measures to their empirically observed values. \textcite{mcdonald-sorensen-turley-2013} fit generalized beta distributions types I and II to income data, with particular eye toward their skew and kurtosis, and \textcite{feng-burkhauser-butler-2006} use the generalized beta type II distribution to correct for censoring in Current Population Survey income data. See \textcite{chotikapanich-etal-2018} for a review of recent literature on the generalized beta type II distribution. \textcite{mcdonald-ransom-2008} fit the generalized beta distribution to income data, and \textcite{bandourian-mcdonald-turley-2002} estimate and numerically evaluate the fit of several subfamilies of the generalized beta distribution.

A second strain in the literature has combined Pareto and log-normal distributions in various ways. \textcite{clementi-gallegati-2005} model incomes using a probability distribution that is log-normal below some cutoff value and Pareto above the cutoff. \textcite{hajargasht-griffiths-2013}, \textcite{reed-2003}, and \textcite{reed-jorgensen-2004} model the income distribution with a product of log-normal and double Pareto distributions, which \textcite{reed-2003} introduced and calls the double Pareto-log-normal distribution. More recent work has focused on taking convex combinations of different probability densities, which are called mixture models. \textcite{schneider-2015} estimates three different models for income: exponential and Pareto, log-normal and Pareto, and all three together. In a similar paper, \textcite{scharfenaker-schneider-2019} model income using a mixture of exponential and Pareto distributions. See \textcite{schneider-scharfenaker-2020} for a review of using mixture distributions to model income. A common theme in the work of Scharfenaker and Schneider on this topic is choosing which probability distributions to mix based on models of labor market segmentation. In a paper that lies somewhere between the second and third strains, \textcite{kaldasch-2012} models the macroeconomy such that capital income follows a log-normal distribution for small values and a Pareto distribution for large values.

Finally, a third strain of this literature has developed out of econophysics, where models typically assume that incomes follow an exponential distribution below some cutoff and a Pareto distribution above the cutoff. \textcite{dragulescu-yakovenko-2003} argue that because money transfers between two economic agents do not change the total amount of money those two agents possess, the amount of money in the economy as a whole should be conserved.\footnote{Of course, monetary theorists will note that issuing or repaying loans changes the total money supply.} Under this assumption and if the economy maximizes entropy, agents will possess different amounts of money that follow an exponential distribution. Following this reasoning, \textcite{derzsy-neda-santos-2012} and \textcite{dragulescu-yakovenko-2001} fit a combination of exponential and Pareto distributions to income data. \textcite{chatterjee-chakrabarti-2007} model low incomes as a gamma distribution arising from a stochastic process with a money conservation rule. See \textcite{shaikh-2020} for a review of the econophysics literature on income.

I fit the models in this paper to publicly available Current Population Survey microdata on income from 1967 through 2023. This is a valuable dataset for analyzing various economic and demographic trends, and it contains a broad measurement of total personal income that includes wages and salary, self-employment income, pension and retirement income, welfare and disability payments, capital income, social security payments, rent income, and alimony. Because the Current Population Survey tracks nominal income, my model describes nominal incomes. I was fortunate to be able to use internal U.S. Census Bureau data files to conduct my analysis. The internal datasets contain less top-coding and other anonymity protections relative to publicly available Current Population Survey microdata files. In principle, top-coding could affect the calculations for this paper, but the results from the internal data look essentially identical to the results from using publicly available data. See the replication files linked on the first page for more information.

Using survey data may present drawbacks in general, but I do not believe it is a problem for this paper. \textcite{bartels-metzing-2019}, \textcite{bricker-etal-2016-aer, bricker-etal-2016-brookings}, and \textcite{burkhauser-etal-2012} have found that surveys generally yield lower estimates of top incomes than tax data. However, surveys often outperform administrative data on low incomes, and \textcite[][p.\ 125]{bartels-metzing-2019} write that away from the tail of the distribution, household surveys provide ``rich information'' about income and demographics. My primary objective is to classify the income distribution as a whole, so having accurate data on a large portion of incomes in the body is more important than having accurate data on the small number of tail incomes. For this reason, using survey data is appropriate, although it means I would need additional data, such as tax data or the Survey of Consumer Finances, to validate my conclusions for annual incomes much greater than \$1 million. Doing so is beyond the scope of this paper, and I leave that calculation for future research.

\heading{2. Choosing a Good Model}

\begin{figure}[p]
\centerline{\bfseries Table \fignum: Distributions Estimated\strut}
\begin{tabularx}{\textwidth}{>{\raggedright\arraybackslash}Xcc}
\toprule
Name & Density & Parameters\\\midrule
Shifted inverse-gamma &
  $\displaystyle y=\frac{\beta^\alpha}{\Gamma(\alpha)}
  \frac{e^{-\frac{\beta}{x-c}}}{(x-c)^{1+\alpha}}$ & 3\\[15pt]
Constant-shift-scale inverse-gamma* &
  $\displaystyle y=\frac{\hat\beta^\alpha}{\Gamma(\alpha)}
  \frac{e^{-\frac{\hat\beta}{x-\hat c}}}{(x-\hat c)^{1+\alpha}}$ & \llap{$\hbox{fewest}\to{}$}1\\[15pt]
Shifted Davis & 
  $\displaystyle y=\frac{\beta^\alpha}{\Gamma(\alpha)\zeta(\alpha)}
  \Bigg(
  \frac1{\raise 0.5ex\hbox{$\big[$}e^{\frac{\beta}{x-c}}-1
    \raise 0.5ex\hbox{$\big]$}(x-c)^{1+\alpha}}
  \Bigg)
  $ & 3\\[12pt]\midrule\noalign{\vskip 2pt}
Shifted generalized beta type II &
  $\displaystyle y=\frac{\alpha(x-c)^{\alpha p-1}}
  {{\textstyle\beta^{\alpha p}}B(p,q)\left(1+\left(\frac{x-c}{\beta}
  \right)^\alpha\right)^{p+q}}$ & 5\\[25pt]
Shifted Dagum & 
  $\displaystyle y=\frac{\alpha p(x-c)^{\alpha p-1}}
    {\textstyle\beta^{\alpha p}
    \left(1+\left(\frac{x-c}\beta\right)^\alpha\right)^{p+1}}$ & 4\\[25pt]
Shifted Burr (Singh-Maddala) &
  $\displaystyle y=\frac{\alpha q(x-c)^{\alpha-1}}
  {\textstyle\beta^\alpha
  \left(1+\left(\frac{x-c}\beta\right)^\alpha\right)^{q+1}}$ & 4\\[25pt]
Shifted Fisk (log-logistic) & 
  $\displaystyle y=\frac{\alpha(x-c)^{\alpha-1}}
  {\textstyle\beta^\alpha
  \left(1+\left(\frac{x-c}\beta\right)^{\!\alpha}\right)^2}$
    & 3\\[20pt]\midrule\noalign{\vskip 5pt}
Shifted log-normal and Pareto (cutoff) & 
  $\displaystyle y=\begin{cases}
  \displaystyle\frac{1}{(x-c)\sigma\sqrt{2\pi}}e^{-(\log(x-c)-\mu)^2/2\sigma^2}
    &\hbox{if $x<k$}\\[\bigskipamount]
  \displaystyle\frac{\alpha x_m^\alpha}{(x-c)^{1+\alpha}}&\hbox{if $x\geq k$}
  \end{cases}$ & 6**\\[35pt]
Shifted log-normal and Pareto (mixture) & \phantom{$y={}$}%
  $\displaystyle \begin{gathered}y=\lambda\left[\frac1{(x-c)\sigma\sqrt{2\pi}}
  e^{-(\log(x-c)-\mu)^2/2\sigma^2}\right]+{}\\
  (1-\lambda)\left[\chi_{x\geq x_m+c}
  \frac{\alpha x_m^\alpha}{(x-c)^{1+\alpha}}\right]\end{gathered}$ & 6\\
\bottomrule
\end{tabularx}
\captionsetup{labelformat=table}
\caption{Table of distributions estimated in the paper. Each distribution is a shifted version of a common probability distribution, where the parameter $c$ controls the shift. Without dimension reduction, the shifted inverse-gamma, Fisk, and Davis distributions have the fewest parameters among models in this paper. As I discuss later, the Davis distribution does not fit as well as the other two, and dimension reduction is inappropriate for the Fisk distribution.\\
\hbox to 1.5em{\hfil}*In the constant-shift-scale inverse-gamma distribution, the $\hat\beta$ and $\hat c$ parameters are functions of $\alpha$ and the year $t$. I abbreviate them here for brevity.\\
\hbox to 1.5em{\hfil}**In the case of a shifted log-normal and Pareto cutoff model, the distribution has 6 parameters but only 5 degrees of freedom because the probability density function needs to integrate to 1.}\label{distributions table}
\end{figure}

In evaluating which probability distributions most closely match income data, we can set baseline requirements according to the qualitative features of U.S. incomes, similar to the principles outlined in \textcite{dagum-1977}. The most apparent empirical characteristics are the following:
\begin{enumerate}\advance\rightskip2em\relax
\item \textbf{Power law tail.} Any model of income should asymptotically follow a power law.
\item \textbf{Single mode.} U.S. incomes consistently show a single mode. The mode occurs relatively close to \$0.
\item \textbf{Positive density at 0.} Plotting sample density of income typically shows histogram bars with positive height at \$0.
\item \textbf{A few negative incomes.} Some small percent of reported incomes are negative.
\end{enumerate}
Most models of income currently in use satisfy points 1 and 2, although we find a few exceptions. Historical choices with a medium or thin tail such as, respectively, log-normal or gamma distributions clearly do not meet the first criterion. The support of the generalized beta or generalized beta type I distributions does not have to extend to infinity.\footnote{But the type II distribution does have support $[0,\infty)$.} We reject any combination of just exponential and Pareto distributions on the grounds that such a combination does not exhibit the necessary mode. \textcite{bandourian-mcdonald-turley-2002} claim that the Weibull distribution fits best among two-parameter distributions they tested, but this model has a thin, exponential tail, not a power law. \textcite{scharfenaker-schneider-2019} incorporate a mixture of exponential and log-normal distributions, which does not have a power-law tail.

Points 3 and 4 are more subtle. It is empirically observable that the income density is positive at \$0---the graph of the 1997 U.S. income distribution in \textcite[][p.\ 11]{bandourian-mcdonald-turley-2002} is a good example. Yet the literature has traditionally modeled incomes with probability distributions whose density at \$0 is 0. Because analysis of the income distribution almost always excludes negative incomes, the result is to favor models that are very steep at \$0.\footnote{\textcite{dagum-1977} notably calls it desirable that a model of the income distribution be able to account for negative incomes. That being said, the support of the Dagum distribution is in fact $[0,\infty)$. \textcite{kleiber-kotz-2003} include almost no discussion of negative incomes beyond Dagum's desideratum, and the current literature has progressed similarly.} Although they lack a proper mode, combinations of exponential and Pareto distributions show positive probability density at \$0 because of the structure of the exponential distribution, and Schneider's (\citeyear{schneider-2015}) mixture of exponential, log-normal, and Pareto distributions satisfies points 1--3 above.

I claim that incorporating negative incomes into the analysis allows us to resolve this mismatch between the data and current modeling choices. Sample density of income data does typically approach 0, just not until somewhere around $-\$10{,}000$. Instead of looking for probability distributions whose support ends at \$0, we should consider probability distributions (1) whose support contains a (small) portion of the negative real line and (2) whose probability density function approaches 0 at the left endpoint of the support. No common probability distribution satisfies these properties, but if we translate common probability density functions left, we end up with models that do. Throughout this paper, I use $c$ to denote a shift parameter, and if $X$ is a random variable, a translated version of $X$ means the random variable $X+c$.\footnote{Adding $c$ to $X$ translates the cumulative distribution function and the probability density function of $X$ by $c$ units to the right. (With negative $c$, both functions move left.) If the support of $X$ is $[0,\infty)$, then the support of $X+c$ is $[c,\infty)$. If the probability density function of $X$ approaches 0 at the left endpoint of its support, then $X+c$ satisfies the two considerations outlined earlier in this paragraph for $c<0$.} I incorporate this additional shift parameter into every distribution I estimate, and intuitively, the shift parameter is the smallest income that should occur in the population with positive chance. Changing the shift parameter is equivalent to adding or subtracting the same amount of money to or from everyone's income.

I estimate shifted versions of models from recent literature that are consistent with the four points above, namely the following:
\begin{itemize}
\item Generalized beta type II, Dagum, Burr, Fisk
\item Log-normal and Pareto mixture
\item Log-normal below a cutoff and Pareto above the cutoff
\end{itemize}
I also estimate a shifted inverse-gamma distribution because it is simple, closely related to the well-known gamma distribution, and satisfies the qualitative requirements for our model. Finally, the historical Davis distribution possesses the same overall appearance as these other models and does not nest any of them as a special case, so I estimate it as well. Table~\ref{distributions table} lists the distributions I estimate in this paper and their probability density functions. Throughout the paper, I use $\beta$ and $c$ to denote scale and shift parameters---most, but not all, other parameters are shape parameters. I compare all these models more closely in section~4, and broadly speaking, all models in this paper fit income data reasonably well.

\heading{3. Parameterizing Income with Inverse-Gamma}

\begin{figure}[p]
\centerline{\bfseries Figure \fignum: Fitted Constant-Shift-Scale Inverse-Gamma Distribution\strut}
\centerline{\includegraphics{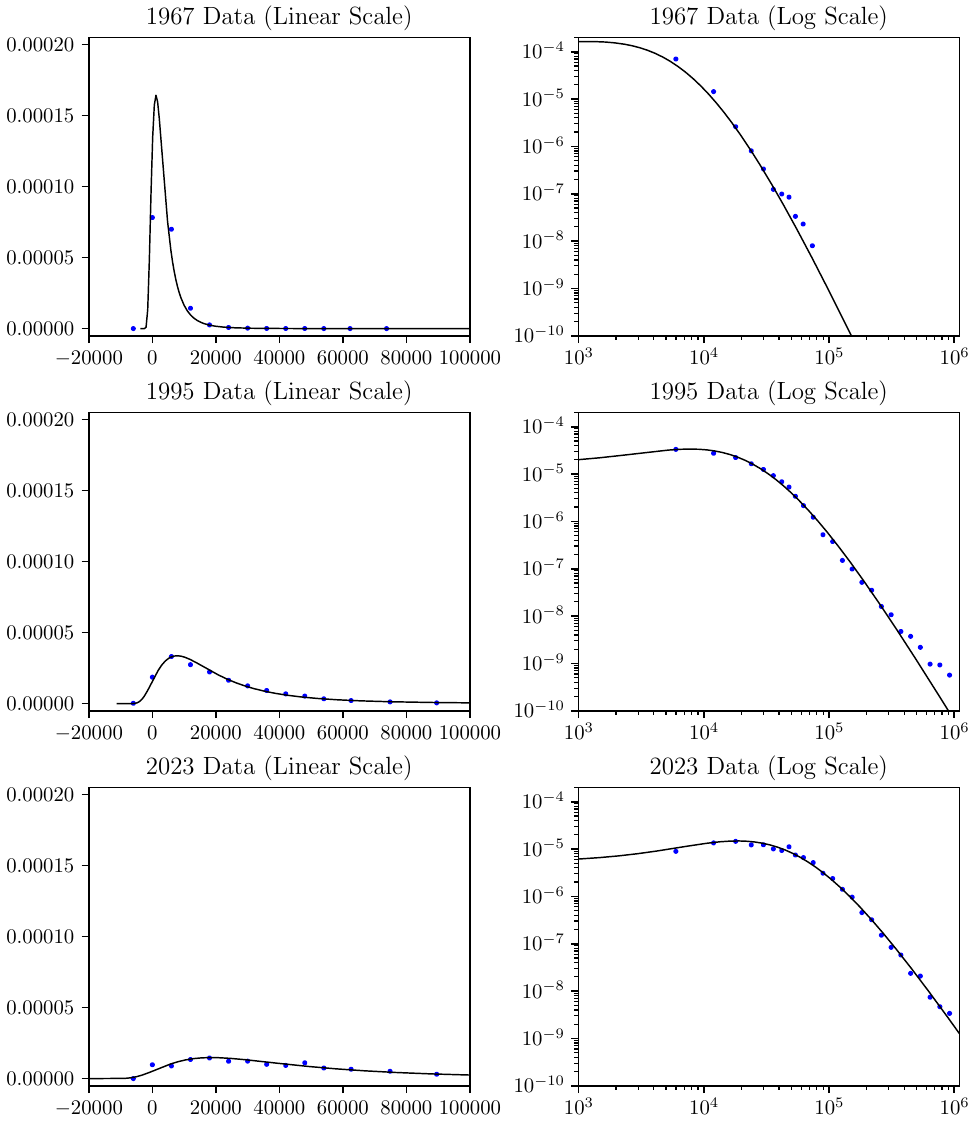}}
\caption{Constant-shift-scale inverse-gamma model fitted to income data in the years 1967, 1995, and 2023. The black curve is the probability density function with estimated parameters, and the blue dots are sample density points. The left column shows graphs on a linear plot, which lets us compare fit in the body of the distribution, and the right column shows graphs on a log-log plot, which is effective for assessing fit in the tail.}\label{densities}
\end{figure}

In this section, I fit Fisk and inverse-gamma distributions to Current Population Survey income microdata, and I demonstrate overfitting and show how to reduce the inverse-gamma distribution to a single degree of freedom. Of the standard probability distributions in this paper, the Davis, Fisk, and inverse-gamma models have the fewest number of parameters. As I discuss in the next section, the Davis distribution does not fit as well as the other two, so we take Fisk and inverse-gamma as baseline quantitive descriptions of the data. For context, an unshifted inverse-gamma distribution is the reciprocal of a gamma distribution, and the logarithm of an unshifted Fisk distribution follows a logistic distribution. If we replace $\beta$ and $c$ in the inverse-gamma distribution by $\hat\beta_t$ and $\hat c_t$, where $\hat\beta_t$ and $\hat c_t$ are linear functions of $\alpha$ and time, we end up with what I have been calling the constant-shift-scale inverse-gamma distribution. I explain why this type of dimension reduction works for the inverse-gamma distribution but not the Fisk distribution later in this section.

Fitting an unshifted Fisk or inverse-gamma distribution to data is straightforward with maximum likelihood, but this approach becomes numerically intractible if we try to estimate $c$ at the same time. (During my exploratory analysis, the shift parameter did not play nicely with likelihood maximization or with other model parameters.) Accordingly, my estimation procedure involves two steps. First, derive a formula for $\alpha$ and $\beta$ conditional on data and $c$ using maximum likelihood, and second, choose the value of $c$ such that the corresponding fitted model minimizes the Kolmogorov-Smirnov statistic. For an empirical cumulative distribution $\hat F$ and model cumulative distribution $F$, the Kolmogorov-Smirnov statistic is given by
\[
\max_{x_i}|\hat F(x_i)-F(x_i)|,
\]
where we maximize over data points $x_i$. The statistic is a common nonparametric measure of how far a model deviates from data and is a useful objective function. The constant-shift-scale model has only a single parameter $\alpha$, so I performed the estimation in one step instead of two by minimzing the Kolmogorov-Smirnov statistic directly.

Figure~\ref{densities} shows the fit to the first, middle, and final years of data using the constant-shift-scale inverse-gamma distribution. The left panels use a linear scale, which illustrates behavior in the body of the distribution, and the right panels use a log-log scale, which allows us to see tail behavior. All four qualitative features of the data from section~2 are apparent in the model, and the fit is good over the entire support of the distribution. Other years show similarly clear matches between model and data, so we have captured the structure of the data using only a single degree of freedom. The corresponding figures with a Fisk or (regular) inverse-gamma distribution are available in the supplemental material and look essentially the same, which supports our use of these two models as a quantitive baseline.

\begin{figure}[t]
\centerline{\bfseries Figure \fignum: Shifted Inverse-Gamma Parameter Estimates\strut}
\centerline{\includegraphics{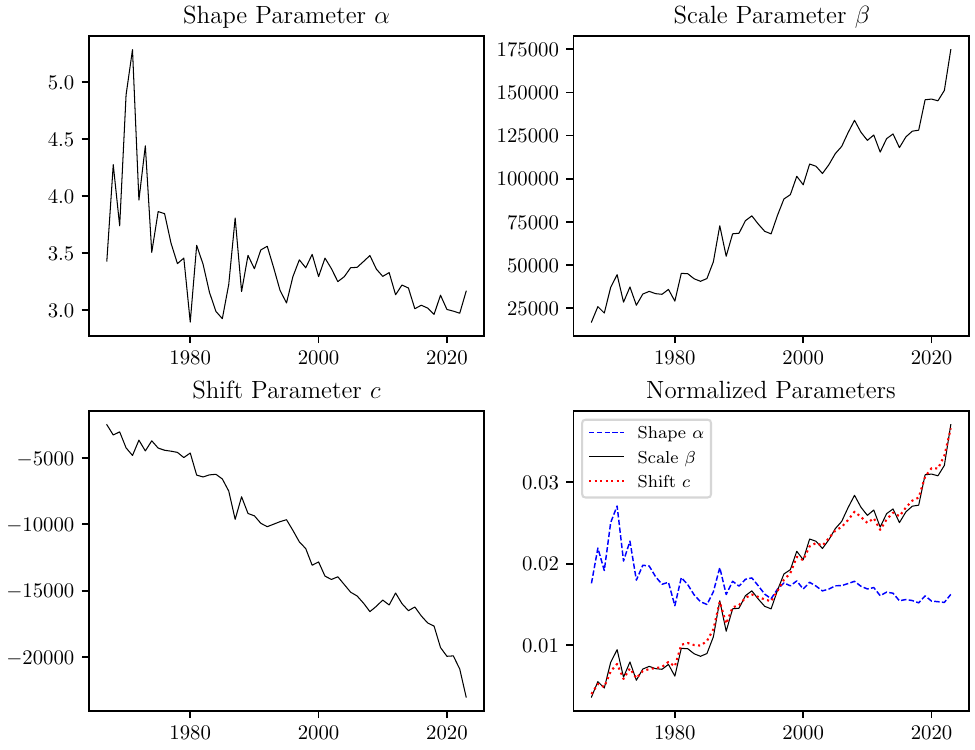}}
\caption{Time series of parameter estimates in each year. The trends are noisy, but we can say that the shape parameter is decreasing, which means more inequality. The estimates change in the same direction from year to year, which suggests a relationship between the underlying parameters. \textbf{Lower right:} Normalized values of the estimates shown in the other three panels. The normalization makes the common dynamics even easier to see.}
\label{parameter-values}
\end{figure}

\begin{figure}[t]
\centerline{\bfseries Figure \fignum: Relationships between Inverse-Gamma Parameters\strut}
\centerline{\includegraphics{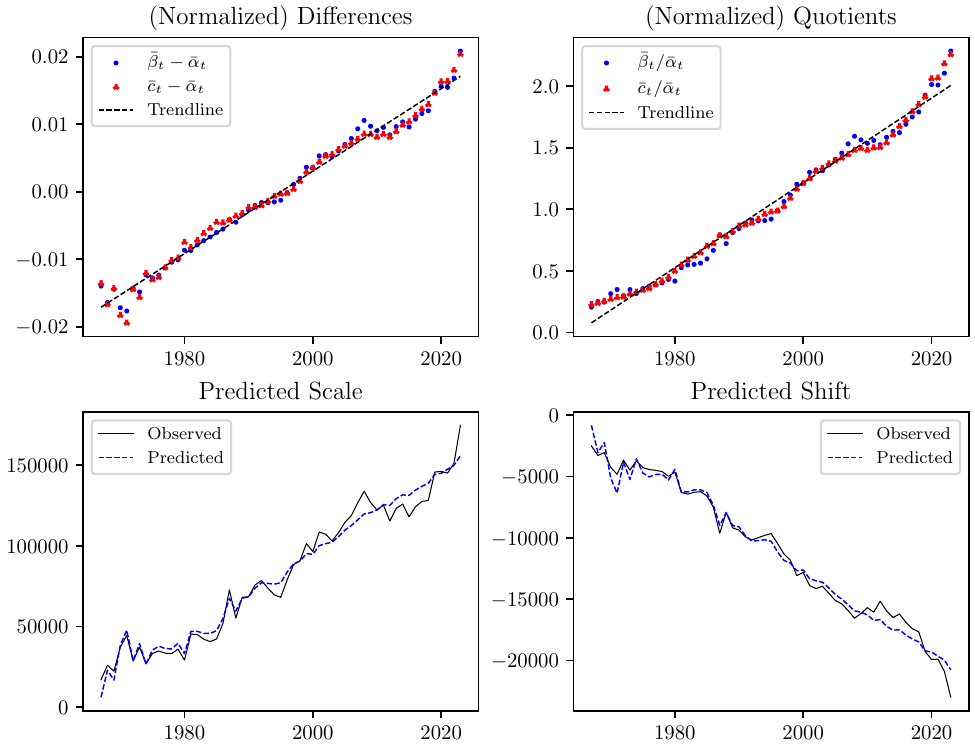}}
\caption{This figure illustrates the linear relationship between time and shifted inverse-gamma parameter estimates. \textbf{Upper left:} The first panel plots differences in normalized parameters from figure~\ref{parameter-values}. The blue dots are the difference $\bar\beta_t-\bar\alpha_t$, and the red trigons are the difference $\bar c_t-\bar\alpha_t$. \textbf{Upper right:} The second panel shows quotients of normalized parameters, using the same notation. Dividing parameters appears to capture as much structure as subtracting, but the inverse-gamma model with proportional parameters does not fit data as well as with linear parameters. \textbf{Lower row:} The third and fourth panels compare the observed and predicted values of the scale and shift parameters respectively under the multilinear regression. The solid black lines are the observed values of parameters, and the dashed blue lines are the predicted values.}
\label{parameter-relationships}
\end{figure}

\begin{figure}[tb]
\centerline{\bfseries Figure \fignum: Normalized Fisk Parameters\strut}
\centerline{\includegraphics{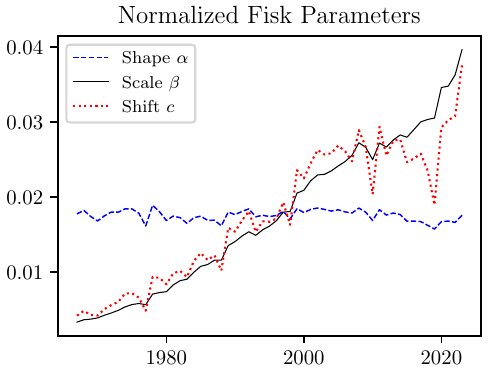}}
\caption{Normalized parameter estimates for the Fisk distribution. Like with the inverse-gamma model, the estimates move in the same directions as each other from year to year, which supports our understanding that the model overparameterizes the data. However, the parameter estimates do not have a clear relationship, so it is not obvious how we would reduce the dimension of this model.}
\label{fisk-normalized}
\end{figure}

Figure~\ref{parameter-values} shows parameter estimates from the shifted inverse-gamma model. The first three panels plot the shape, scale, and shift parameter estimates. During the past 50 years, the shape and shift parameters are decreasing, and the scale parameter is increasing. Because lower shape parameter means fatter tail, a decreasing shape parameter points to increasing income equality in the last half-century. Also notable are the similarities in all three time series. The fourth panel illustrates this phenomenon by graphing normalized estimates $\bar\alpha_t$, $\bar\beta_t$, and $\bar c_t$ on the same plot, where
\begin{align*}
\bar\alpha_t&=\frac{\alpha_t}{\sum_{\text{years}\ i}\alpha_i}&
\bar\beta_t&=\frac{\beta_t}{\sum_{\text{years}\ i}\beta_i}&
\bar c_t&=\frac{c_t}{\sum_{\text{years}\ i}c_i}.
\end{align*}
The normalized estimates have the same magnitude, so it is easy to compare behavior. Because $\bar\beta_t$ and $\bar c_t$ are practically on top of one another, we conclude that $\beta\propto c$. It is also clear that $\bar\alpha_t$ has a time-dependent relationship with $\bar\beta_t$ and $\bar c_t$. All three normalized parameters move in the same direction from year to year, and the difference between trends is monotonic with respect to time.

The last panel in figure~\ref{parameter-values} raises the possibility of a linear relationship between parameters. The difference between the trends for $\bar\alpha_t$ and the other two parameters decreases steadily as we move right along the $x$-axis. The upper left panel of figure~\ref{parameter-relationships} plots these differences, and they follow line across time periods. The other possibility for a simple relationship between parameters is one where they are all proportional, and the upper right panel of figure~\ref{parameter-relationships} plots quotients of normalized parameters to explore this possibility. The quotients also lie on a line, so proportionality seems plausible. The changes in $\bar\alpha_t$ are much larger than the changes in $\bar\beta_t$ and $\bar c_t$ in the earliest years of data, and the pattern gradually reverses as we move forward in time, which we should expect to see in such a scenario. However, an inverse-gamma distribution with a proportional relationship between parameters fits early years of data poorly, so I reject this possibility. See the supplemental material for more information. Figure~\ref{fisk-normalized} shows normalized parameters for the Fisk distribution. The estimates move together like with inverse-gamma, which suggests that the Fisk distribution also overparameterizes income data. However, the parameters do not have an obvious relationship to one another, so it is not clear if or how we could reduce the dimension like with inverse-gamma.

\begin{figure}[t]
\centerline{\bfseries Table \fignum: Constants for the Constant-Shift-Scale Model\strut}
\centerline{\begin{tabularx}\textwidth{lCCCC}\toprule
Constant & $\phi$ & $\psi_0$ & $\psi_1$ & $\psi_2$\\
Value & $-0.133$ & $\$727\mathord,800$ & $-\$366/\text{year}$ & $-\$2194$ \\\bottomrule
\end{tabularx}}
\captionsetup{labelformat=table}
\caption{Coefficients that quantify the linear relationship between parameter estimates under a shifted inverse-gamma model of income. The value of $\phi$ is the constant of proportionality between $\beta$ and $c$. In determining $\beta$ and $c$, $\psi_0$ is a constant term. The $\psi_1$ coefficient is the time effect, and $\psi_2$ is the shape effect.}\label{CSS_InvG_constants}
\end{figure}

\begin{figure}[t]
\centerline{\bfseries Figure \fignum: Constant-Shift-Scale Inverse-Gamma Parameter Estimates\strut}
\centerline{\includegraphics{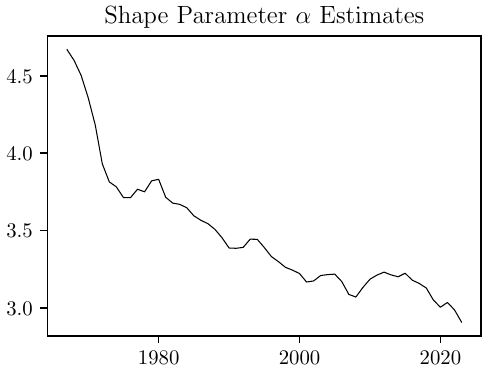}}
\caption{Time series of the parameter estimate in each year for a constant-shift-scale inverse-gamma model of income. The single (shape) parameter looks similar to the shape parameter under a shifted inverse-gamma model, but it is smoother, which suggests it is capturing less noise in the data.}
\label{CSS_InvG_parameters}
\end{figure}

We quantify the relationship between these parameters by modeling them as
\begin{equation}
c_t=\phi\beta_t=\psi_0+\psi_1t+\psi_2\alpha_t
\label{parameter-equation}
\end{equation}
and choosing $\phi$ and $\psi_i$ appropriately.\footnote{A proportional relationship means having $c_t=\phi\beta_t=\alpha_t(\psi_0+\psi_1t)$ instead of equation (\ref{parameter-equation}), so all three parameters would be proportional to one another, rather than just $\beta$ and $c$ being proportional. As I discuss in the supplemental material, this model performs worse than equation (\ref{parameter-equation}).} If $\beta$ and $c$ are proportional, the constant of proportionality should be the ratio of their magnitudes, so
\[
\phi=\sum_tc_t\bigg/\sum_t\beta_t,
\]
Typically linear regression minimizes the distance between a one-dimensional output variable and one or more columns of data, but if we want to base our choice of $\psi_i$ on both $\beta$ and $c$, we need to manage two output variables simultaneously. I approached the fitting problem by adding together separate penalties for the scale and shift parameters. If $A$ is a matrix containing a column of ones, a column of years and one column for values $\alpha_t$, our objective function $L$ is
\[
L=||C-A\psi||^2+||\phi B- A\psi||^2,
\]
where $\psi=(\psi_0,\psi_1,\psi_2)\T$ is the vector of $\psi_i$ values, and $B$ and $C$ are vectors of parameter estimates for $\beta$ and $c$ in each year. As is standard, we use the $L^2$ norm, which means we are taking a least-squares approach. Because $\phi$ is the constant of proportionality between $B$ and $C$, the objective function penalizes deviations in either term equally, and applying linear algebra and some multivariable calculus, we get
\begin{equation}\label{phi and psi}
\psi=\frac12(A\T A)^{-1}A\T(\phi B+C).
\end{equation}
The lower panels of figure~\ref{parameter-relationships} show predicted values for $\beta_t$ and $c_t$ under this multilinear regression compared with the original estimates. The predicted values match the observed values.

Table~\ref{CSS_InvG_constants} shows the results of equation (\ref{phi and psi}). The shift parameter has opposite sign to and is about one eighth the magnitude of the scale parameter. The starting year number in our calendar determines $\psi_0$, so this constant is largely uninteresting. The $\psi_1$ and $\psi_2$ constants quantify how much the shift parameter $c$ changes when the year and the shape parameter respectively increase by 1, and dividing by $\phi$ gives the corresponding effects on $\beta$. It follows that we can decompose any change to the income scale into a time effect and a shape effect as
\[
\beta_{t_1}-\beta_{t_0}=\frac{\psi_1}{\phi}(t_1-t_0)+
  \frac{\psi_2}\phi(\alpha_{t_1}-\alpha_{t_0})
\]
Over the last 60 years, the scale parameter has risen by about \$150,000, and the shape parameter has decreased by 1.5. The shape effect for this change is roughly $-\$25{,}000$, so we can attribute the growth in income scale entirely to the natural spreading of the distribution. Changes in the income distribution's shape have served as a drag on the scale.

Building the linear relationship into the inverse-gamma density function gives us the constant-shift-scale inverse-gamma distribution. We replace $\beta$ and $c$ with their functions of $\alpha$ and $t$, which gives us
\begin{equation}
y=\frac{(\psi_0+\psi_1t+\psi_2\alpha)^\alpha}{\phi^\alpha\Gamma(\alpha)}
  \frac{e^{-\frac1\phi\frac{\psi_0+\psi_1t+\psi_2\alpha}
  {x-(\psi_0+\psi_1t+\psi_2\alpha)}}}{(x-(\psi_0+\psi_1t+\psi_2\alpha))^{1+\alpha}},
\label{CSS_InvG_density}
\end{equation}
where $t$ is the year. Figure~\ref{CSS_InvG_parameters} shows estimates of $\alpha$ for all years of data, and bootstrapping based on \textcite{jolliffe-2003} indicates that standard errors are about 1\% of parameter values. The trend in $\alpha$ is similar to that from the regular inverse-gamma model, but the time series is smoother, which suggests that this model captures less noise. I suspect that some of the parameter behavior in the earliest years may be due to lack of data because later years of the Current Population Survey have much better coverage of large incomes than the earliest datasets. Equation~(\ref{CSS_InvG_density}) has more variables than shifted inverse-gamma density, but $\alpha$ is the only degree of freedom because $t$ is exogenous and $\phi$, $\psi_i$ are the same across datasets. To be technically precise, we went from 3 parameters per year of data to 1 parameter per year of data plus 4 time-independent constants.

\heading{4. Comparison with Other Distributions}

\begin{figure}[p]
\centerline{\bfseries Figure \fignum: Fits of Other Models to 2023 Income Data\strut}
\centerline{\includegraphics[width=6.5in]{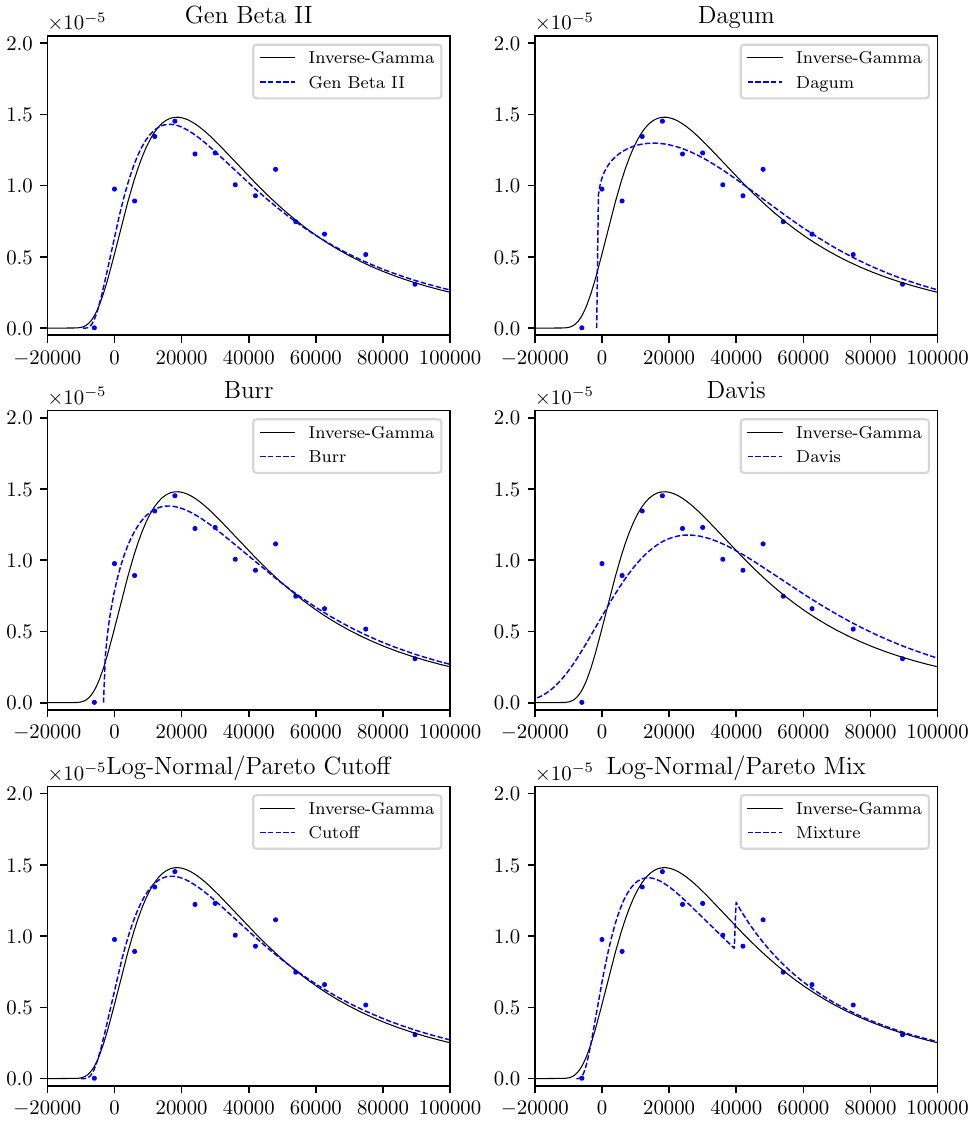}}
\caption{Comparison between constant-shift-scale inverse-gamma distribution and other statistical models of income using 2023 income data, specifically the generalized beta type II, Dagum, Burr, Davis, and log-normal and Pareto combinations. The solid black line is the fitted constant-shift-scale inverse-gamma model, and the dashed blue line is the other model of interest. The blue dots are sample density points. In all cases, with the possible exception of the Davis and Dagum distributions, the fit is good---no model obviously fails to match the body of the data.}
\label{linear-comparison}
\end{figure}

\begin{figure}[p]
\centerline{\bfseries Figure \fignum: Fits of Other Models to 2023 Income Data\strut}
\centerline{\includegraphics[width=6.5in]{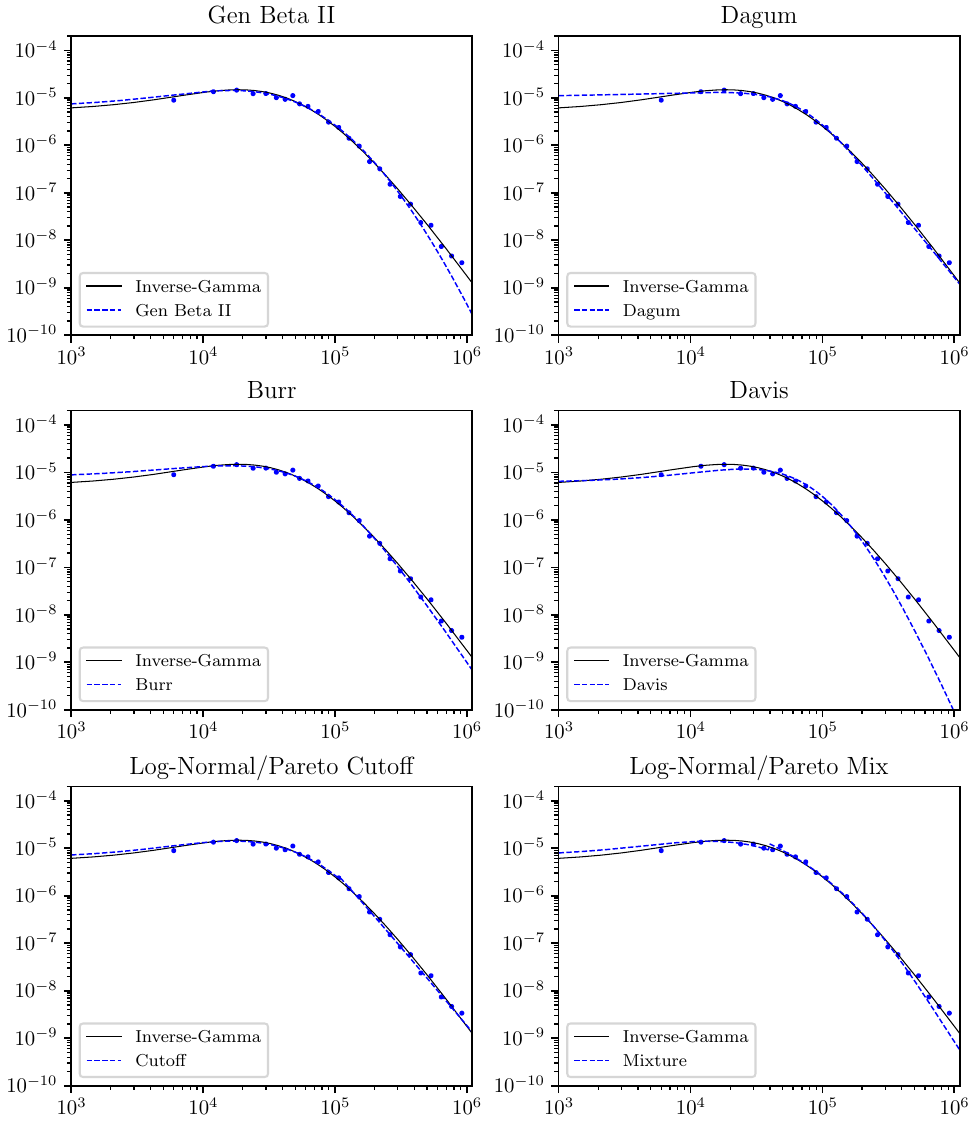}}
\caption{Comparison of the same models as in Figure~\ref{linear-comparison}, except with a focus on the tails. Again, the solid black line is the constant-shift-scale inverse-gamma model, and the dashed blue line is the model of interest. The blue dots are sample density points. Because we selected only models with a power-law tail, the fit in the tails is good in all cases, although the Davis distribution drops off faster than the other models around the largest incomes.}\label{loglog-comparison}
\end{figure}

In this section, I compare results from section~3 to models from the literature, and I find that constant-shift-scale inverse-gamma distribution fits at least as well as any other model. The generalized beta type II distribution is a particular favorite in the literature because of its flexibility and good performance with data. For specific values of parameters, this model becomes equal to Dagum (when $q=1$), Burr (when $p=1$), and Fisk (when $p=q=1$) distributions. The Davis distribution has received less attention in recent years, and it resembles an inverse-gamma distribution with extra mass in the tail. The log-normal and Pareto cutoff and mixture models combine log-normal and Pareto density functions by either using a cutoff or taking a convex combination, respectively. All models that I estimate in this paper have support $[c,\infty)$ and a tail that (asymptotically) follows a power law.

I estimated parameters the same way as for the Fisk and inverse-gamma distributions: (1) find the maximum likelihood estimator conditional on $c$, data, and possibly one or two other parameters; then (2) find the value of missing parameters that minimizes the Kolmogorov-Smirnov statistic. For generalized beta type II, Dagum, Burr, Davis, and log-normal and Pareto mixture distributions, I maximized the likelihood function numerically.\footnote{Maximizing the generalized beta type II and log-normal and Pareto mixture likelihood functions was a surprisingly challenging numerical problem because very similar probability density functions can correspond to very different parameters. Nelder-Mead is a particularly stable algorithm and the only algorithm I tried that gave meaningful answers. When they estimate generalized beta type II parameters, \textcite{graf-nedyalkova-2014} use the method-of-moments estimators for the Fisk distribution as their initial guess for the algorithm, and changing the guess even slightly can make the algorithm diverge.} The maximum likelihood conditions for the log-normal and Pareto cutoff model reduce to a single equation, which makes estimation relatively easy. Finding the correct $c$-value for the Dagum, Burr, Davis, and cutoff models was numerically straightforward, but the other two distributions required a more complicated approach. See the appendix for details.

Figures~\ref{linear-comparison} and \ref{loglog-comparison} show the estimation results for 2023 income data compared with a constant-shift-scale inverse-gamma model. Figure~\ref{linear-comparison} uses a linear scale, and figure~\ref{loglog-comparison} uses a log-log scale. In all plots, the fitted inverse-gamma distribution is the solid black curve, and the dashed blue curve is the other model of interest. Several observations are apparent. First, in all cases, the fit is good---no model obviously fails to match the data. Second, the Burr and Dagum distributions approach their left endpoints very steeply. The boundary of the support is the smallest income we expect to see with any significant probability, and the Dagum and Burr models set this value too high. Third, the Davis distribution puts too little mass around the mode. Fourth, the log-normal and Pareto cutoff model has a small kink, which we see in the log-log plot around $\$120{,}000$, and no discontinuity while the mixture model has a visible discontinuity around $\$40{,}000$ at the left endpoint of the Pareto density. These facts point to at best limited gains in fit from using one of the models here instead of inverse-gamma.

At the same time, any successful model should use the smallest possible number of parameters, and as we saw in the previous section, the minimum number of parameters for U.S. incomes is 1. When a model has too many parameters, it can easily capture noise instead of signal if small changes in data cause large fluctuations in parameter estimates. The (regular) shifted inverse-gamma distribution overfits the data, so even as few as 3 parameters is still too many. In this section, the effect is more extreme because with the higher-dimensional models, large swaths of parameter space produce nearly identical behavior. For the generalized beta type II and the log-normal and Pareto mixture distributions, the problem is severe enough to raise identifiability concerns. I conclude that models in the current literature have too much flexibility, and accordingly, I offer the constant-shift-scale inverse-gamma distribution as a model that fits income data without overparameterizing.

\heading{5. Interpretation and Applications}

Besides the obvious ability to include it in larger models of the macroeconomy, a parametric description of the income distribution provides a number of advantages for understanding income. First, knowing the mathematical determinants of the distribution can point us in the right direction toward figuring out why incomes change in the aggregate. We observe two types of changes in the data: a linear time-dependent scaling effect and a nonlinear dynamic reshaping effect. The linear relationship between parameters and time holds across all years of data, so any economic factors that affect short-term income dynamics must enter the model through the shape parameter $\alpha$. My analysis did not yield any quantities that perfectly correlate with $\alpha$, and I leave it as an open question what $\alpha$ means in practical terms. An answer to this question will tell us exactly what causes the distribution to evolve over time, apart from the linear scaling.

Second, we can interpret the parameters and discuss relationships between them. The scale parameter $\beta$ tracks how incomes as a whole grow multiplicatively, so adjusting $\beta$ for inflation converts the model between describing real and describing nominal incomes. The shift parameter $c$ corresponds to the smallest amount of income that anyone receives with any significant probability, and changing $c$ is equivalent to adding or subtracting an amount of money to everybody's income. The literature does not ascribe particular meaning to shape parameters, but mathematically, we can consdier the shape parameter to be an ordinal measure of income inequality. Lower shape parameter means a fatter tail and therefore higher inequality.

One of the most intriguing results from this paper is the linear relationship between parameters. The shift parameter is a constant multiple of the scale parameter, so all changes to income scaling perfectly account for any changes to the smallest income that we expect to see in the population. It is a question for future research why $c/\beta=\phi$, which is the left endpoint of the support when we normalize the distribution, has remained constant at a negative number. Further, the shape parameter and thus the amount of inequality has a linear relationship to the scale parameter. More investigation is needed to determine the mechanism at play, but we can say observationally that higher inequality goes hand-in-hand with a smaller income scale. Since we normally expect independence between parameters, this relationship presents an opportunity for future research on determinants of inequality. As the linear relationship cannot continue indefinitely into the past, it is another question for future research how far back it goes and what happened previously.

Third, the linear scaling of nominal incomes with respect to time presents new insights for understanding individual income dynamics. If the shape parameter is constant at a level $\alpha$ over multiple years, the scale parameter $\beta_t$ increases by $\psi_1/\phi$ per year, and the shift parameter decreases by $\psi_1$ per year. These changes to parameters dilate the probability density function around the origin, which suggests that individual incomes may follow a diffusion process, perhaps with discrete jumps to account for a new job or change in life circumstances. Since $\alpha$ directly affects the shift and scale parameters, any diffusion process will interact with whatever factors determine $\alpha$, and the overall dynamics may be complex. I leave it as a question for future research how this diffusion process could work and what it says about humans that we may organize our society this way.

We can apply the same concept to percent changes in annual income. Suppose $x$ represents some income, and we run a thought experiment where we scale $x$ by an amount corresponding to the additional time-effect from $\beta_{t}$ to $\beta_{t+1}$. If the shape parameter $\alpha$ stays constant from time $t$ to $t+1$, the time-effect is the only change in $\beta_t$, and the percent change in $x$ is
\begin{align*}
\frac{\frac{\beta_{t+1}}{\beta_t}x-x}{x}&=\frac{\beta_{t+1}}{\beta_t}-1\\
&=\frac{\beta_{t}+\psi_1}{\beta_t}-1\\
&=\frac{\psi_1}{\beta_t}.
\end{align*}
As we increase $t$, $\beta_t$ gets larger, so the percent change in $x$ between $t$ and $t+1$ gets smaller. If we want a measure that stays constant as we evolve incomes without changing the distribution's shape, we should multiply the percent difference by $\beta_t$. Compare this model with a situation where income grows exponentially with time, for example $\beta_t=\beta_0e^{\psi_1t}f(\alpha_t)$. In this case, the percent change in income is
\begin{align*}
\frac{\frac{\beta_{t+1}}{\beta_t}x-x}{x}&=\frac{e^{\psi_1}\beta_t}{\beta_t}-1\\
&=e^{\psi_1}-1,
\end{align*}
which remains constant as we move through time. My analysis suggests that because nominal income scales linearly with respect to time and not exponentially, we should expect percent income changes from year to year to decrease in absence of other effects, albeit rather slowly.

The linear relationship between $\beta$ and $t$ also means that absent changes in the shape parameter, per-capita income increases linearly with respect to time. Per-capita income is the expected value of the distribution, so under a constant-shift-scale model, per-capita income in year $t$ is
\begin{equation}\label{per-capita income}
\frac{\beta_t}{\alpha_t-1}+c_t=c_t\left(\frac1\phi\frac1{\alpha_t-1}+1\right).
\end{equation}
If $\alpha_t$ is constant at level $\alpha$ over a number of years, then equation (\ref{per-capita income}) increases by 
\[
\psi_1\left(\frac1\phi\frac1{\alpha-1}+1\right)
\]
per year. Multiplying equation (\ref{per-capita income}) by population gives total income in a given year, and if the total income increases exponentially while $\alpha_t$ stays constant, which may happen for example in a steady state of some dynamical system macroeconomic model, the exponential growth in total income must be due to population dynamics, not to changes in the income distribution. When $\alpha_t$ does change from year to year, the picture becomes more complicated, but unless $\alpha$ changes very fast or approaches 1, we should expect equation (2) to remain essentially linear in time.

\begin{figure}[t]
\centerline{\bfseries Figure \fignum: Understanding the Gini Coefficient\strut}
\centerline{\includegraphics[width=6.5in]{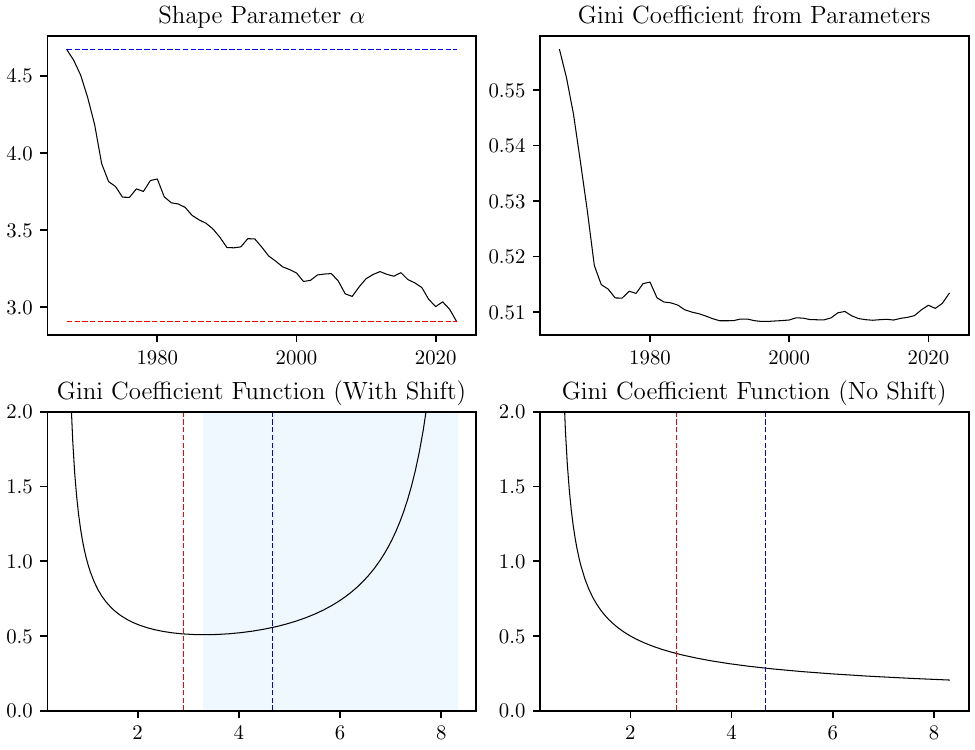}}
\caption{This figure explores the Gini coefficient under a constant-shift-scale inverse-gamma income distribution. \textbf{Upper left:} The first panel repeats the information from Figure~\ref{CSS_InvG_parameters} for reference.\textbf{Upper right:} The second panel is the Gini coefficient that we calculate using equation (\ref{gini coefficient function}). \textbf{Lower left:} The third panel plots equation (\ref{gini coefficient function}), where the pale blue region corresponds to the portion of the real line where the Gini coefficient becomes non-monotonic in $\alpha$. \textbf{Lower right:} The fourth panel shows the analogue of equation (\ref{gini coefficient function}) for an unshifted inverse-gamma distribution. In all cases, the dashed lines indicate maximum and minimum parameter estimates.}\label{gini}
\end{figure}

Fourth, we can achieve deeper insight by linking formulas to the economic interpretation of parameters than we can through non-parametric analysis alone. For example, consider inequality measures. The literature on income uses a variety of statistics to quantify inequality, and I consider the Gini coefficient here as one of the most popular.\footnote{The literature on income inequality's relationship to other social and economic issues is enormous, and it encompasses topics such as financial regulation and policy \autocite{bumann-lensink-2016, christopoulos-mcadam-2017, de-haan-sturm-2017, ghossoub-reed-2017, kirschenmann-malinen-nyberg-2016}, development \autocite{chiu-lee-2019, huang-etal-2015, meniago-asongu-2018}, wealth inequality \autocite{benhabib-bisin-luo-2017}, health \autocite{curran-mahutga-2018, detollenaere-etal-2018, pickett-wilkinson-2015, vincens-emmelin-stafstrom-2018}, environmental factors \autocite{liu-zheng-wang-2020}, and covid-19 outcomes \autocite{elgar-stefaniak-wohl-2020, liao-de-maio-2021, oronce-etal-2020, tan-etal-2021}.} See \textcite{de-maio-2007} for a glossary of inequality measures. Under a constant-shift-scale inverse-gamma distribution, the Gini coefficient comes out to
\begin{equation}\label{gini coefficient function}
\frac{1}{1+(\alpha-1)\phi}
\frac{\Gamma\left(\alpha-\frac12\right)}{\sqrt\pi\Gamma(\alpha)}.
\end{equation}
We know that lower $\alpha$ means a fatter tail and higher inequality, and for an unshifted inverse-gamma distribution, the Gini coefficient is a monotonic transformation of $\alpha$. However, when some incomes are negative, the Gini coefficient becomes non-monotonic and can be greater than 1, which makes no sense, and equation (\ref{gini coefficient function}) becomes infinite when $\alpha=1-1/\phi\approx 8.5$, which suggests a breakdown in the model for $\alpha$ near $1-1/\phi$. Mathematically, the problem is that when incomes can be negative, the notion of total income loses the same intuitive meaning it has when all incomes are positive.

Figure~\ref{gini} illustrates this point. The first two panels show the values of $\alpha$ and corresponding Gini coefficient value, but even though inequality increases, the Gini coefficient decreases, suggesting lower inequality. The third panel explains why. This panel shows the graph of equation (\ref{gini coefficient function}), and under a constant-shift-scale inverse-gamma distribution of income, the Gini coefficient is very non-monotonic. When $\alpha$ is below about 3.5, the Gini coefficient decreases with $\alpha$, but for greater $\alpha$, the Gini coefficient increases despite lower inequality and eventually approaches a singularity at $1-1/\phi$. It follows that using $\alpha$ as an ordinal measure of inequality may make more sense than the Gini coefficient or any other measure that is unsuited to negative incomes. For discussion of negative incomes and their role in measuring inequality, see \textcite{hlasny-ceriani-verme-2021, raffinetti-siletti-vernizzi-2015}.

Fifth, a parametric description of the income distribution makes it easy to check predictions from other models. Proportional random growth models, for example, have a long history of being used to model power laws in economics, although \textcite{gabaix-etal-2016} question whether a proportional random growth mechanism can fully explain recent income dynamics. In a proportional random growth model, all agents start out with an initial income, and in each time period $t$, we multiply agent $i$'s income by a random variable $X_{i,t}$, where all $X_{i,t}$ are independent and identically distributed, have nonnegative support, and satisfy some regularity conditions. The central limit theorem guarantees that the distribution of agent incomes converges to a log-normal distribution, and if we exogenously impose a minimum income, the process converges to a Pareto distribution instead. Other enhancements such as incorporating births and deaths exist in the literature, and see \textcite{gabaix-2009} for a review of power laws in economics, including proportional random growth models.

We can use our parameterization to evaluate whether a given dynamic framework results in an accurate income distribution---simply simulate the model, and see if the result resembles a shifted inverse-gamma distribution. Unfortunately, I do not know of any proportional random growth process that aggregates correctly in the case of income. One problem is that matching the distribution of negative incomes is tricky when dynamics are multiplicative and independent of the agent's current income level. A second difficulty is that we want any random growth model to capture the linear relationship between parameters and time, and it is not obvious how to do that. More generally, if a model does not produce the correct distribution when we simulate it, we may be able to adjust model assumptions based on differences between the simulated data and the target distribution.

Sixth, we can draw conclusions from the probability distribution itself. Two avenues for future research come immediately to mind in this regard. First, the inverse-gamma distribution has a natural interpretation as a rate. If $\alpha$ is an integer, then the (unshifted) inverse-gamma distribution is the distribution of arrival rates of the first $\alpha$ events in a homogeneous Poisson process with rate parameter $\beta$. Shifting right by $c$ units complicates the picture, but because income is also a rate, specifically a rate of dollar accumulation, I view this connection as a potentially fruitful avenue for understanding why U.S. income data exhibits such clear structure. 

The other immediate avenue for future research comes from the fact that conditional on $c$, $X$ is a shifted inverse-gamma random variable if and only if $X$ maximizes entropy subject to the constraints
\begin{align}\label{maxent}
\E\log(X-c)&=\log(\beta)-\psi(\alpha)&
\E\left(\frac1{X-c}\right)&=\frac\beta\alpha
\end{align}
Maximum-entropy probability distributions are ubiquitous in physics and chemistry and intuitively correspond to the ``most random'' arrangement of the world that is possible under some stated conditions. It remains an open question whether models with maximum entropy are appropriate to describe the macroeconomy, but if they are appropriate, then the constraints in equation (\ref{maxent}) completely determine the shape of the income distribution. No other model that I estimate in this paper is a maximum-entropy probability distribution in a meaningful sense. For more information on entropy maximization, see \textcite{callen-1985} and \textcite{guenault-1995}.

\heading{7. Conclusion}

I began this paper with four qualitative observations about income data and a commitment to fitting parameteric models that match those observations. Adding a shift parameter in models from the literature improves their fit to U.S. income data. We do not have a clear winner on fit alone, and dimensionality concerns suggest that we should use a low-dimensional model. The shifted Fisk and inverse-gamma distributions have 3 parameters each and show signs of overfitting, and identifiability issues crop up with some of the higher-dimensional models in this paper. It is not obvious how we could reduce the dimension of the Fisk distribution, but the inverse-gamma parameter estimates are linearly related to one another. This relationship indicates that the inverse-gamma distribution captures fundamental structure in our data and allows us to construct a one-dimensional model of income, which I offer for use in future work on this topic. I highlight the possible diffusion process at work, the interpretation of an inverse-gamma distribution as a rate, and the connection to entropy maximization as promising avenues for future research.

\bigskip\medskip
\centerline{\bfseries References}

{
\def\mkbibnamefamily#1{\textsc{#1}}
\def\mkbibnamegiven#1{\textsc{#1}}
\def\mkbibnameprefix#1{\textsc{#1}}
\def\mkbibnamesuffix#1{\textsc{#1}}

\printbibliography[heading=none]

\vskip-\lastskip

}

\vfill\eject

\heading{Appendix: Estimation Procedure}

Three aspects of the data and estimation are worth discussing in detail. First is the data preparation. Incomes of \$0 are extremely overrepresented in the data, to the point where it would technically be more accurate to represent the income distribution as an inverse-gamma density with an atom at \$0. This atom is not the signal I am trying to capture, so I dropped all entries with income of \$0. Doing so does not affect the density at \$0 because probability density incorporates information across an interval, not just at a single income. This means my statistical model says nothing about null incomes, but because my primary interest is people who actually receive income, that is okay. To prevent any issues with outliers, I also excluded from consideration a handful of observations on either end of the support that fell above or below some maximum or minimum cutoff.

The second aspect is the two steps of the parameter estimation. In the first step, I calculated maximum-likelihood estimators conditional on the shift parameter $c$ and possibly one or two other parameters. See table~\ref{estimation-procedures} for details. In all equations that follow, $n$ is the number of data points, and we assume that the survey weights $w_i$ are normalized to sum to $n$. At a number of places in the estimation, I drew on technical insights from \textcite{chen-2016, dunster-2006, graf-nedyalkova-2022}. The shifted inverse-gamma and shifted log-normal and Pareto cutoff distributions have almost-closed-form maximum likelihood estimators. For the shifted inverse-gamma distribution, the maximum-likelihood estimate for $\alpha$ solves the equation
\[
\log\alpha-\psi(\alpha)-\log\left(\frac1n\sum_{i=1}^n\frac{w_i}{x_i-c}\right)
  -\frac1n\sum_{i=1}^nw_i\log(x_i-c)=0.
\]
The corresponding value of $\beta$ is
\[
\beta=\frac\alpha{\displaystyle\frac1n\sum_{i=1}^n\frac{w_i}{x_i-c}}.
\]
The equation for $\alpha$ is numerically well behaved, so we can solve it easily, which gives us $\alpha$ and $\beta$ in terms of $c$ and data.

For the cutoff model, we take $k$ and $c$ as given. We maximize the likelihood function subject to the constraint that the probability density function integrates to 1. Doing so gives $\alpha$ as
\[
\alpha=\frac1{\displaystyle\frac1{n^\uparrow}\sum_{i=j+1}^nw_i\log\left(\frac{x_i-c}k\right)},
\]
where $n^\uparrow$ is the total survey weight on observations with incomes above $k$, and $j$ is the index that divides incomes below $k$ from incomes above $k$. (Similarly, $n_\downarrow$ is the total survey weight on observations with incomes below $k$.) The log-normal parameter $\mu$ solves
\[
\sum_{i=1}^jw_i\log(x_i-c)-n_\downarrow\mu+n^\uparrow\sqrt{p_0+p_1\mu}
  \;\frac{\displaystyle\Phi'\left(\frac{\log k-\mu}{\sqrt{p_0+p_1\mu}}\right)}
    {\displaystyle 1-\Phi\left(\frac{\log k-\mu}{\sqrt{p_0+p_1\mu}}\right)}=0
\]
where $\Phi$ is the cumulative distribution function of the standard normal distribution, and $p_0$ and $p_1$ are given by
\begin{align*}
p_0&=\frac1{n_\downarrow}\sum_{i=1}^jw_i\log(x_i-c)^2
  -\frac{\log k}{n_\downarrow}\sum_{i=1}^jw_i\log(x_i-c)\\
p_1&=\log k-\frac1{n_\downarrow}\sum_{i=1}^jw_i\log(x_i-c).
\end{align*}
Thus we can find $\mu$ numerically. At the same time, $\sigma^2=p_0+p_1\mu$, and the mass condition on the density gives
\[
x_m=k\left[1-\Phi\left(\frac{\log k-\mu}{\sigma}\right)\right]^{1/\alpha}
\]
for $x_m$. Altogether these equations give us $\mu$, $\sigma^2$, $\alpha$, and $x_m$ in terms of $c$, $k$, and data.

\begin{figure}[t]
\centerline{\bfseries Table \fignum: Estimation Procedure Overview\strut}
\begin{tabularx}{\textwidth}{>{\raggedright\arraybackslash}Xll}\toprule
Name & Maxized Likelihood Using & Parameters Remaining\\\midrule
Constant-shift-scale inverse-gamma & N/A & Numerically find $\alpha$\\
Shifted inverse-gamma & Reduces to single equation & Numerically find $c$\\
Shifted Davis & Two-step numerical & Numerically find $c$\\
Shifted generalized beta type~II & Two-step numerical & Brute force find $c$\\
Shifted Dagum & Nelder-Mead & Numerically find $c$\\
Shifted Burr & Nelder-Mead & Numerically find $c$\\
Shifted Fisk & Nelder-Mead & Numerically find $c$\\
Shifted log-normal and Pareto (cutoff) & Reduces to single equation & Numerically find $c$, $k$\\
Shifted log-normal and Pareto (mixture) & Three-step numerical & Brute force find $c$, $x_m$\\\bottomrule
\end{tabularx}
\captionsetup{labelformat=table}
\caption{Overview of the estimation procedure for each probability distribution. The general procedure was (1) find maximum-likelihood estimator conditional on some parameters including the shift parameter and (2) then find the remaining parameters such that the fitted model minimizes an objective function. Although no model has a purely closed-form maximum likelihood estimator, ``reduces to a single equation'' means that the maximum likelihood equations result in a single equation of one variable that we can solve numerically. ``Nelder-Mead'' means that I used a Nelder-Mead algorithm to maximize the likelihood function, and multi-step numerical means that I applied various analytical and numerical techniques in two or three steps.}\label{estimation-procedures}
\end{figure}

For the remaining distributions, we cannot find a closed form (or almost-closed form) for the maximum-likelihood estimator. For Dagum, Burr, and Fisk distributions, I maximized the likelihood function directly using a Nelder-Mead algorithm. The proper choice of initial guess is crucial, and \textcite{graf-nedyalkova-2014} recommend using the method-of-moment estimators for the Fisk distribution as the initial guess for $\alpha$ and $\beta$:
\begin{align*}
\alpha&=\frac\pi{\sqrt{3v}}\\
\beta&=e^m
\end{align*}
where
\begin{align*}
m&=\frac1n\sum_{i=1}^nw_i\log(x_i)&
v&=\frac1n\sum_{i=1}^nw_i(\log(x_i)-m)^2
\end{align*}
The authors also recommend setting $p=q=1$ as appropriate. This method worked well, but slightly changing the initial guess or changing the numerical maximizer usually caused the algorithm to fail, which indicates high sensitivity to the estimation procedure.

For the Davis distribution, the maximum-likelihood equations are
\begin{gather*}
\psi(\alpha)+\frac{\zeta'(\alpha)}
  {\zeta(\alpha)}+\frac1n\sum_{i=1}^n w_i\log(x_i-c)=\log\beta\\
\frac1n\sum_{i=1}^n
  \left(\frac{w_i}{x_i-c}\right)\frac{e^{\frac\beta{x_i-c}}}
  {e^{\frac\beta{x_i-c}}-1}=\frac{\alpha}{\beta}
\end{gather*}
I found $\alpha$ and $\beta$ in two steps. I first calculated $\alpha$ conditional on $\beta$ and $c$ and then picked the value of $\beta$ that satisfied the maximum-likelihood conditions.

The generalized beta type II and log-normal and Pareto mixture models are more complicated. For the generalized beta type II distribution, I first found $p$ and $q$ conditional on data and remaining parameters by numerically solving the equations
\begin{align*}
\psi(p+q)-\psi(p)&=\frac1n\sum_{i=1}^nw_i
 \log\left(1+\left(\frac{x_i-c}\beta\right)^{\!\alpha}\:\right)-
 \frac\alpha n\sum_{i=1}^nw_i\log\left(\frac{x_i-c}\beta\right)\\
\psi(p+q)-\psi(q)&=\frac1n\sum_{i=1}^nw_i
 \log\left(1+\left(\frac{x_i-c}\beta\right)^{\!\alpha}\:\right),
\end{align*}
which come from the maximum-likelihood conditions. Then I chose $\alpha$ and $\beta$ numerically to maximize the likelihood using the corresponding $p$ and $q$ values. 

For the log-normal and Pareto mixture model, I first used a truncated maximum-likelihood estimator to find $\mu$ and $\sigma^2$ conditional on other parameters. Let $\eta$ be the fraction of survey weight on points less than $x_m+c$. Then $\mu$ solves the quadratic equation $x^2+bx+c=0$ with
\begin{align*}
b&=\Phi^{-1}\left(\frac\eta\gamma\right)^{\!2}
 \Big[\overline{\log x_i}-\log(x_m)\Big]-2\log(x_m)\\
c&=\Phi^{-1}\left(\frac\eta\gamma\right)^{\!2}
 \left[\log(x_m)\,\overline{\log x_i}-\overline{\log x_i^2}\right]+\log(x_m)^2,
\end{align*}
where 
\begin{align*}
\overline{\log x_i}&=\frac1n\sum_{i=1}^n\log x_i &
\overline{\log x_i^2}&=\frac1n\sum_{i=1}^n\log(x_i)^2.
\end{align*}
Once we know $\mu$, we can find $\sigma^2$ according to
\[
\sigma=\frac1{\textstyle\Phi^{-1}\left(\frac\eta\gamma\right)}(\log(x_m)-\mu),
\]
where $\Phi$ is the cumulative distribution function of the standard normal distribution. After estimating $\mu$ and $\sigma^2$, I numerically maximized the likelihood function to find $\alpha$, and then I found $\gamma$ by numerically solving a maximum likelihood condition for $\gamma$. Specifically, if $l_i$ is the likelihood ratio
\[
l_i=\frac{\alpha x_m^\alpha}{(x_i-c)^{1+\alpha}}\sigma\sqrt{2\pi}
  e^{-(\log(x_i-c)-\mu)^2/2\sigma^2},
\]
then the maximum likelihood estimate for $\gamma$ is a root of
\[
\frac{n_\downarrow}\gamma+\sum_{i=j+1}^n\frac{w_i(1-l_i)}{\gamma+l_i},
\]
where again $n_\downarrow$ is the total survey weight less than $x_m+c$, and $j$ is the index that divides incomes less than $x_m+c$ and incomes greater than $x_m+c$. Finding a numerical root of this expression gave me $\gamma$ in terms of $c$ and $x_m$.

These procedures give us the non-shift parameters in terms of $c$ (and possibly $x_m$ or $k$) and data. For the remaining parameters, I minimized the Kolmogorov-Smirnov statistic. I usually did so numerically, but for the generalized beta type II and mixture models, I implemented a brute-force search because numerical minimizing failed. Large swaths of parameter space produce similar enough behavior that practical identifiability is a concern for these two models.

Finally, to calculate sample density points, I partitioned the support of the data into bins $B_i=[b_i,b_{i+1}]$. I used linear bins for incomes up to \$60,000, and for incomes above \$60,000, I used logarithmic bins. Linear bins are a useful default binning method, and logarithmic bins are standard when dealing with power-law data \autocite{lin-newberry-2023, milojevic-2010, newman-2005}. Each linear bin has a width of $\$6000$, and the ratio of bin endpoints for logarithmic bins is at least 1.2. If $\eta_i$ is the fraction of total survey weight on incomes that fall into $B_i$, then the sample density $s_i$ is given by $\eta_i/(b_{i+1}-b_i)$.

\vfill\eject
\hrule height \z@
\vfill

\heading{Supplemental Material}

\vfill\eject

\begin{figure}[p]
\centerline{\bfseries Figure \fignum: Fitted Inverse-Gamma Distribution\strut}
\centerline{\includegraphics{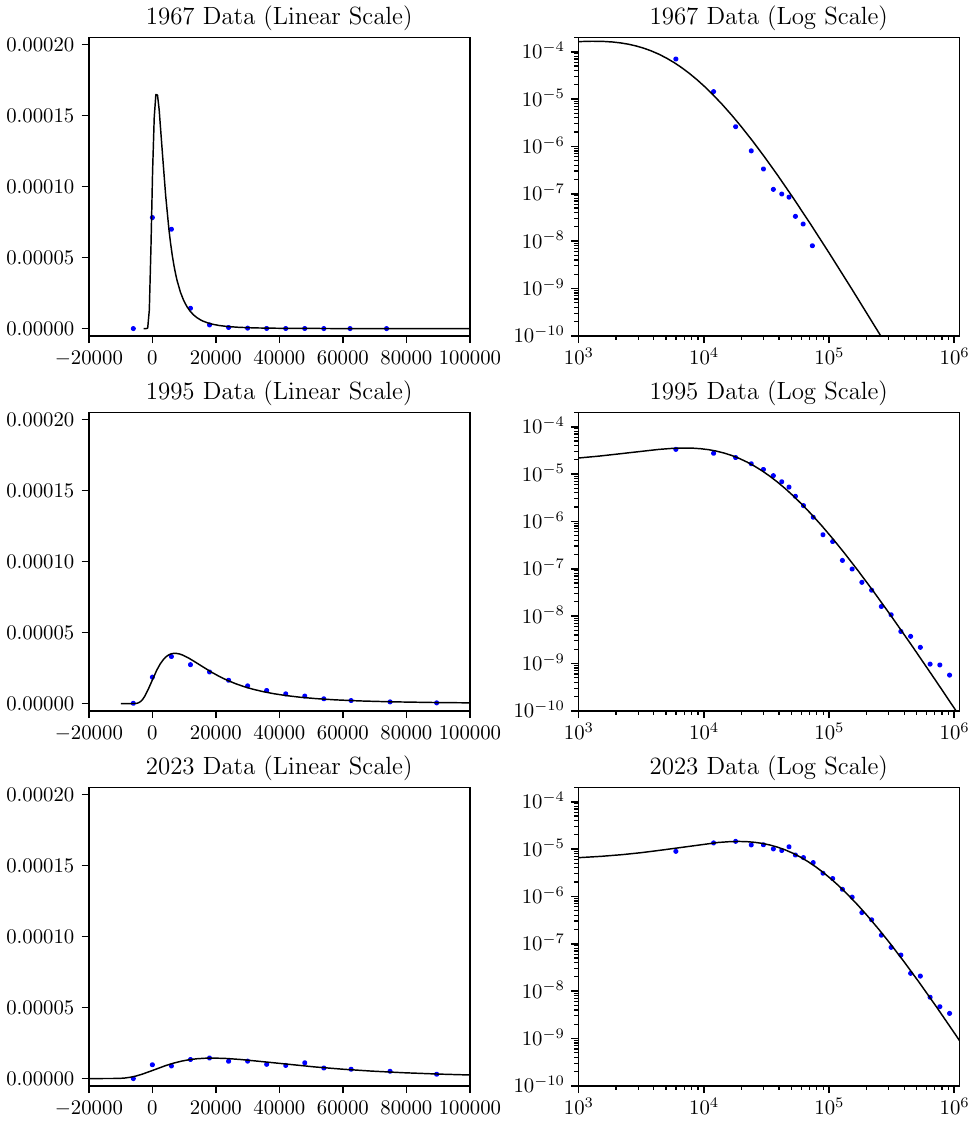}}
\caption{Shifted inverse-gamma model fitted to income data in the years 1967, 1995, and 2023. The black curve is the probability density function with estimated parameters, and the blue dots are sample density points.}
\end{figure}

\begin{figure}[p]
\centerline{\bfseries Figure \fignum: Fitted Fisk Distribution\strut}
\centerline{\includegraphics{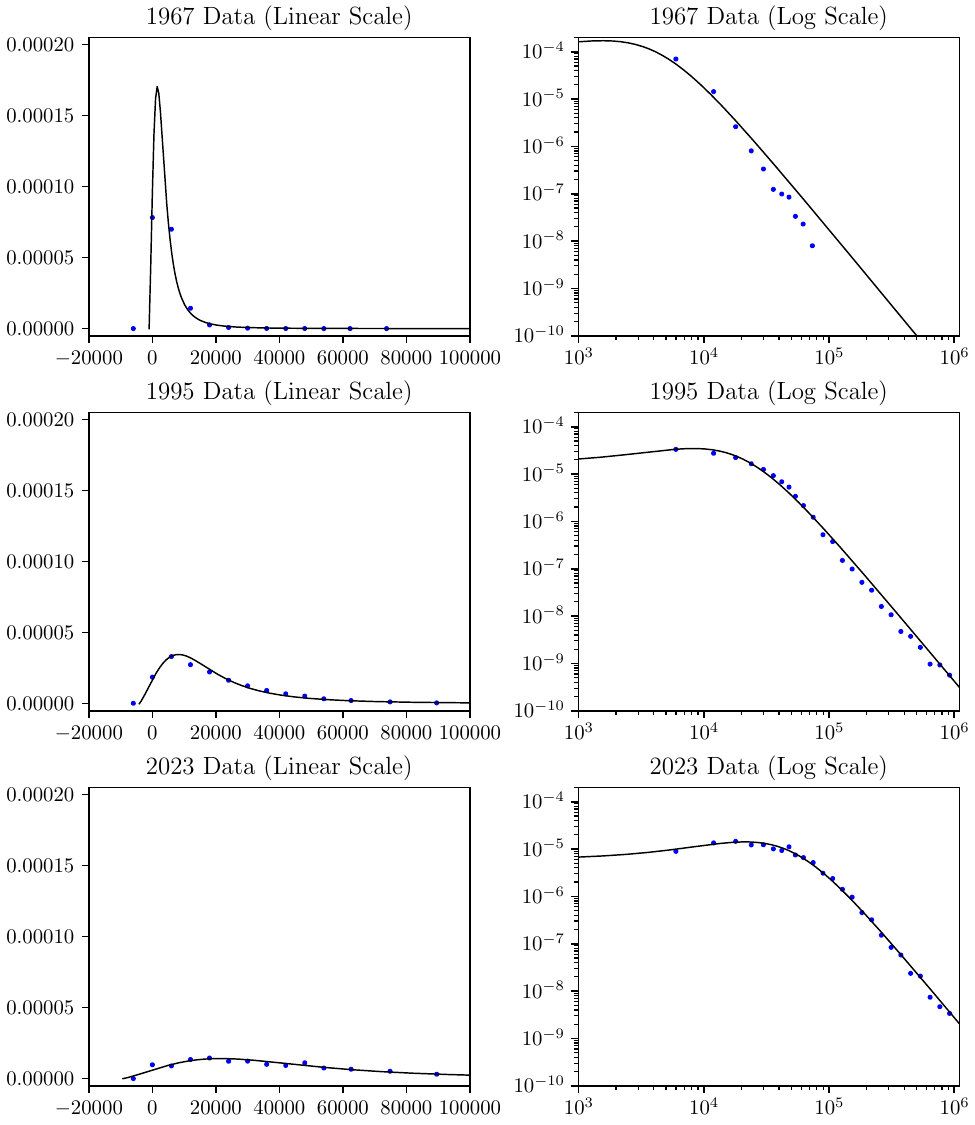}}
\caption{Shifted Fisk distribution fitted to income data in the years 1967, 1995, and 2023. The black curve is the probability density function with estimated parameters, and the blue dots are sample density points.}
\end{figure}

\begin{figure}[p]
\centerline{\bfseries Figure \fignum: Fitted Inverse-Gamma Distribution with Proportional Relationship\strut}
\centerline{\includegraphics{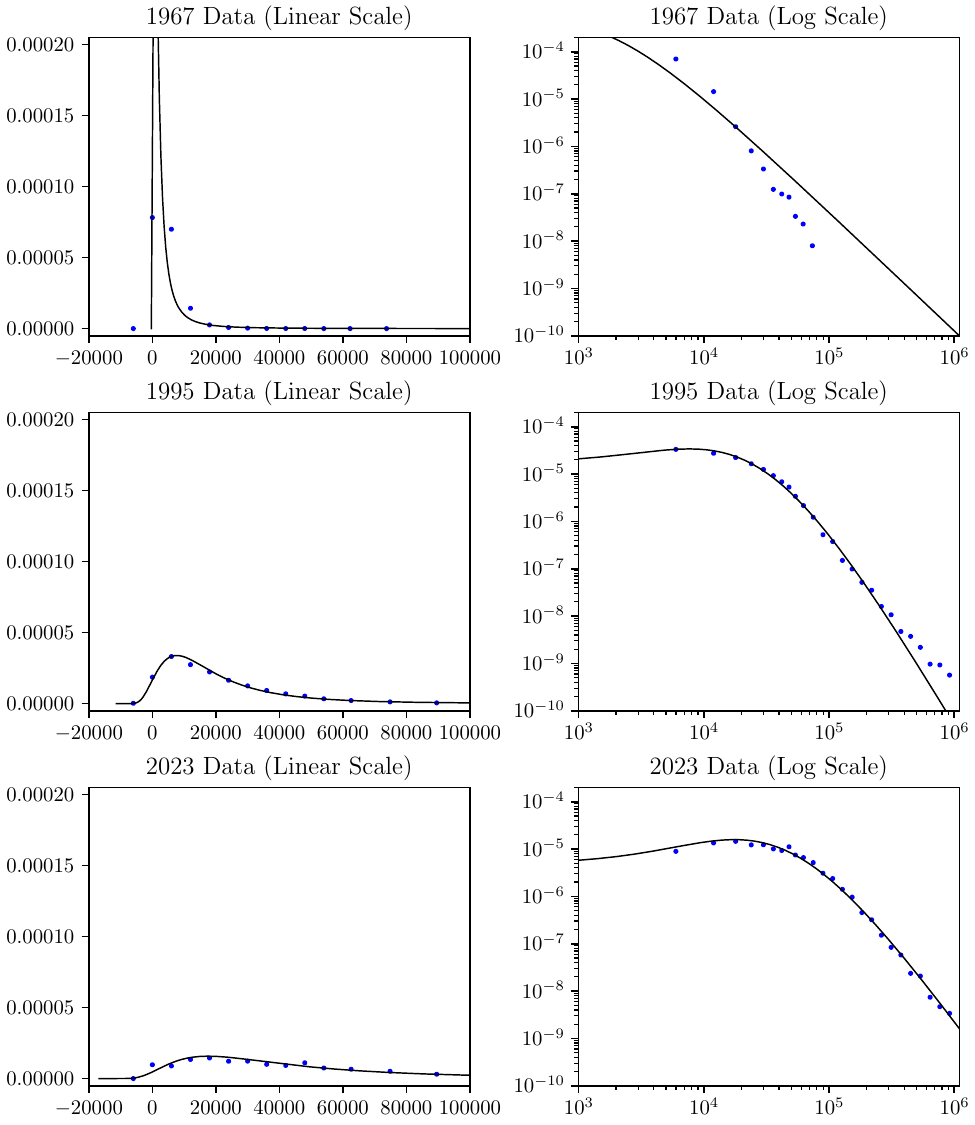}}
\caption{Shifted inverse-gamma distribution fitted to income data in the years 1967, 1995, and 2023, where we impose a proportional relationship between all three parameters. The black curve is the probability density function with estimated parameters, and the blue dots are sample density points. Formally, we impose $c=\phi\beta=\alpha(\psi_0+\psi_1t)$. The fit to 1967 data is poor.}
\end{figure}

\end{document}